\documentclass{article}
\usepackage[a4paper, left=1.5cm, right=1.5cm, top=2.5cm, bottom=2.5cm]{geometry}
\usepackage{graphicx} 
\usepackage{braket}
\usepackage{amsmath}
\usepackage{amssymb}
\usepackage{xcolor}
\usepackage{authblk} 
\usepackage{cite}

\title{Non-Hermitian Quantum Mechanics with Applications to Gravity}

\author[]{Oem Trivedi\thanks{Email: oem.trivedi@vanderbilt.edu}}
\author[]{Alfredo Gurrola\thanks{Email: alfredo.gurrola@vanderbilt.edu}}
\author[]{Robert J. Scherrer\thanks{Email: robert.scherrer@vanderbilt.edu}}

\affil[]{Department of Physics and Astronomy, Vanderbilt University, Nashville, TN 37235, USA}

\date{\today}

\begin{document}

\maketitle

\begin{abstract}
Hermiticity is usually treated as a foundational axiom of quantum mechanics, guaranteeing real spectra and unitary time evolution. In this work we argue that Hermiticity is more naturally understood as a symmetry law as a consequence of the global conservation of an inner-product current. We show that in spacetimes admitting complete Cauchy surfaces without boundary flux, this conservation reduces to the familiar condition $H^\dagger=H$ in the canonical inner product. However, in the presence of causal horizons, most strikingly in black hole geometries—this conservation law is obstructed for restricted observers. Tracing over inaccessible degrees of freedom inevitably yields completely positive trace-preserving dynamics with an effective non-Hermitian generator. Using quantum thermodynamics and the monotonicity of relative entropy, we demonstrate that the generalized second law may be reinterpreted as an entropy balance compensating precisely for the flux of inner-product charge through the horizon. The structure of Einstein’s equations, via the Bianchi identity and the Raychaudhuri focusing equation, provides the geometric mechanism underlying this balance. We also show that Black Hole Ringdown could be a realistic observational probe for this idea, with it also being an avenue to provide quantitative upper bounds on the the strength of horizon-induced inner-product flux. In this way, gravity, entropy production and effective non-Hermiticity are unified under a single structural principle, with Hermiticity emerging as the special case of globally conserved inner-product symmetry.
\end{abstract}

\section{Introduction}
Hermiticity has long been viewed as one of the central structural postulates of quantum mechanics. In the standard Hilbert space formulation, the requirement that physical observables be represented by Hermitian operators guarantees real eigenvalues, ensures unitary time evolution for closed systems and secures the probabilistic interpretation through the Born rule and conservation of the canonical inner product \cite{qm1bohm2013quantum,qm2zettili2009quantum,qm3sakurai2020modern,qm4griffiths2018introduction,qm5shankar2012principles,qm6scherrer2024quantum}. Within this framework, Hermiticity is tightly linked with spectral reality, probability conservation and measurement theory. Over the past several decades, however, non-Hermitian quantum mechanics has evolved into a sophisticated and experimentally accessible domain driven largely by the recognition that many realistic quantum systems are effectively open, with degrees of freedom either deliberately eliminated or fundamentally inaccessible \cite{nh1moiseyev2011non,nh2ashida2020non,nh3hatano1996localization}. In such circumstances complex potentials, gain–loss mechanisms and non-unitary generators arise naturally as effective descriptions. These structures have been implemented and probed in a variety of engineered platforms, where controlled amplification, absorption, and post-selected dynamics allow laboratory access to regimes beyond strictly unitary evolution. Consequently, non-Hermitian models are now studied both as effective theories of open-system dynamics and, in some proposals, as generalized frameworks with modified inner products and biorthogonal constructions that restore probabilistic consistency in a broader sense \cite{nh4jones2014relativistic,nh5gopalakrishnan2021entanglement,nh6hatano1997vortex,nh7bender2007making,nh8longhi2010optical}. This expanding program has been accompanied by precision experimental efforts aimed at constraining effective non-Hermitian generators through interferometric tests, spectral response analyses, decay-profile measurements and consistency checks of norm conservation in appropriately defined inner products \cite{nh9jones2010non,nh10krejvcivrik2015pseudospectra,nh11cui2012geometric,nh12bergholtz2021exceptional}.
\\
\\
On the other hand of the spectrum seems gravitational physics which deals with the dynamics of spacetime itself and with phenomena far removed, at least superficially, from laboratory realizations of non-unitary quantum evolution. Over the last century, general relativity has provided a remarkably successful geometric description of gravitation, explaining planetary motion, gravitational waves \cite{gw1thorne1995gravitational,gw2sathyaprakash2009physics,gw3cai2017gravitational,gw4lasky2015gravitational} and the large scale structure of the universe \cite{gp1will1982theory,gp2nojiri2003new,gp3marolf2009unitarity}. Modern cosmology has revealed a universe dominated by dark energy and dark matter, whose physical origins remain among the deepest open problems in fundamental physics \cite{de1li2013dark,de2copeland2006dynamics,de3peebles2003cosmological,de4mortonson2013dark,dm1bertone2018history,dm2garrett2011dark,dm3bertone2018new,dm4hui2021wave,dm52017dark}. Black hole physics has uncovered a profound interplay between geometry, thermodynamics, and quantum theory culminating in the discovery of Hawking radiation and the formulation of black hole thermodynamics \cite{bh1carlip2014black,bh2witten2025introduction,bh3wall2018survey,bh4davies1978thermodynamics,bh5ong2022black}. At the same time, extensive efforts toward quantum gravity ranging from semiclassical quantum field theory in curved spacetime to string theory, loop quantum gravity and effective field theory approaches have sought to reconcile quantum mechanics with dynamical spacetime. These developments have led to strong puzzles such as the black hole information problem and to refined notions of entropy, horizon dynamics and generalized second laws that intertwine geometry with quantum information \cite{bi1giddings1995black,bi2jacobson2013boundary,bi3danielsson1993quantum,bi4mathur2009information,bi5susskind1997black}.
\\
\\
Could it be possible, then, that gravity and Hermiticity have some fundamental relation? While such a connection may at first appear unlikely, there are reasons to suspect that it lies within the realm of logical possibility. Gravitational physics introduces causal horizons and restricted observer algebras, naturally leading to reduced descriptions that resemble open quantum systems. If Hermiticity is tied to the conservation of a global inner product, then the presence of horizons and boundary fluxes may obstruct that conservation for restricted observers, potentially giving rise to effective non-Hermiticity in a manner dictated not by phenomenology but by spacetime structure itself. In this work, we explore precisely this possibility by examining Hermiticity through the lens of covariant inner-product conservation, quantum thermodynamics, and black hole entropy balance. The work is hence structured as follows. In section 2 we shall discuss a brief overview of non-Hermitian quantum mechanics, while in section 3 we discuss about quantum thermodynamics and its need near Black Holes. In section 4 we establish why non-Hermiticity becomes inevitable near black holes, while in section 5 we introduce Hermiticity as a Symmetry law. In section 6 we discuss the relations of all this with Einstein's theory in terms of structure and we conclude our work in section 7.
\\

\section{Brief Overview of Non-Hermitian Quantum Mechanics}
A minimal extension of standard quantum mechanics arises when one relaxes the requirement that the Hamiltonian be self-adjoint with respect to the canonical Hilbert space inner product, while preserving the Schr\"odinger form of time evolution,
\begin{equation}
i\hbar \frac{\partial}{\partial t}\ket{\psi(t)}=\hat H \ket{\psi(t)}
\label{eq:NH_Schro}
\end{equation}
In ordinary quantum mechanics, Hermiticity of $\hat H$ ensures real spectra and unitary evolution and allowing $\hat H\neq \hat H^\dagger$ modifies the dynamical structure at its root. It is convenient to decompose a general non-Hermitian Hamiltonian into Hermitian and anti-Hermitian parts,
\begin{equation}
\hat H=\hat H_{\rm H}+i\hat \Gamma
\label{eq:H_decomp}
\end{equation}
where $\hat H_{\rm H}=\hat H_{\rm H}^{\dagger}$ and $\hat \Gamma=\hat \Gamma^{\dagger}$. This decomposition is always possible because any operator can be uniquely written as the sum of its Hermitian and anti-Hermitian components and the operator $\hat H_{\rm H}$ governs coherent oscillatory evolution, while $\hat \Gamma$ encodes dissipative or amplifying behavior.
\\
\\
To see this explicitly, consider the time derivative of the norm under the canonical inner product as using Eq.\eqref{eq:NH_Schro} and its adjoint, one finds
\begin{equation}
\begin{split}
\frac{d}{dt}\braket{\psi(t)|\psi(t)}
&=\bra{\dot\psi}\psi\rangle+\bra{\psi}\dot\psi\rangle \\
&=\frac{i}{\hbar}\bra{\psi(t)}\big(\hat H^{\dagger}-\hat H\big)\ket{\psi(t)} \\
&=-\frac{2}{\hbar}\bra{\psi(t)}\hat \Gamma\ket{\psi(t)}
\end{split}
\label{eq:norm_evol}
\end{equation}
Thus, unless $\hat \Gamma=0$, the norm is not conserved and positive expectation values of $\hat \Gamma$ correspond to exponential decay, while negative expectation values lead to amplification. In this sense, $\hat \Gamma$ acts as a gain–loss generator in the reduced or effective description. This behavior is already familiar from effective descriptions of unstable states, optical potentials and resonance phenomena.
\\
\\
The spectral properties of a non-Hermitian Hamiltonian differ qualitatively from the Hermitian case and for this, consider right eigenstates defined by
\begin{equation}
\hat H\ket{n_{\rm R}}=E_n\ket{n_{\rm R}}
\end{equation}
with complex eigenvalues $E_n=E_n^{\rm R}+iE_n^{\rm I}$, then the corresponding time evolution of a stationary state yields
\begin{equation}
e^{-iE_n t/\hbar}=e^{-iE_n^{\rm R}t/\hbar}\,e^{E_n^{\rm I}t/\hbar}
\end{equation}
so that the imaginary part of the spectrum governs decay or growth rates. Because $\hat H$ and $\hat H^\dagger$ no longer share eigenvectors, the spectral problem becomes biorthogonal and one introduces left and right eigenstates satisfying
\begin{equation}
\hat H\ket{n_{\rm R}}=E_n\ket{n_{\rm R}},\qquad
\hat H^{\dagger}\ket{n_{\rm L}}=E_n^{\ast}\ket{n_{\rm L}},\qquad
\braket{n_{\rm L}|m_{\rm R}}=\delta_{nm}
\label{eq:biorth}
\end{equation}
Completeness then takes the form $\sum_n \ket{n_{\rm R}}\bra{n_{\rm L}}=\mathbb{I}$. Observables are naturally evaluated using both left and right states with expectation values written as $\langle \hat O\rangle=\bra{\psi_{\rm L}}\hat O\ket{\psi_{\rm R}}$. This biorthogonal structure ensures internal consistency of the spectral decomposition even when the eigenvalues are complex and eigenvectors are not orthogonal in the usual sense.
\\
\\
A distinguished subclass of non-Hermitian theories permits a generalized notion of unitarity as well, for this consider that there exist a positive definite metric operator $\eta$ such that
\begin{equation}
\hat H^{\dagger}=\eta \hat H \eta^{-1}
\end{equation}
then $\hat H$ is said to be pseudo-Hermitian and in this case, defining a modified inner product $\langle\psi|\phi\rangle_{\eta}=\langle\psi|\eta|\phi\rangle$ restores norm conservation under time evolution. Indeed, one may verify that
\begin{equation}
\frac{d}{dt}\langle\psi(t)|\psi(t)\rangle_{\eta}=0
\end{equation}
provided $\eta$ is time independent. This construction underlies $\mathcal{PT}$-symmetric quantum mechanics and related frameworks in which non-Hermitian Hamiltonians with real spectra admit a consistent probabilistic interpretation. In such cases, apparent non-Hermiticity with respect to the canonical inner product reflects a different Hilbert space geometry rather than a breakdown of quantum consistency.
\\
\\
Beyond spectral considerations, non-Hermitian quantum mechanics has found wide application in effective theories of open systems, non-equilibrium statistical mechanics and engineered gain–loss platforms as well. Complex potentials arise naturally in optical systems, cold atom setups, and condensed matter contexts where amplification and dissipation can be tuned experimentally. Exceptional points, non-Hermitian phase transitions, and sensitivity enhancement near spectral degeneracies have become central themes of current research. From a dynamical standpoint, non-Hermitian terms may be viewed as encoding coarse-graining, measurement backaction, or environmental conditioning providing a compact parametrization of deviations from strict unitarity \cite{nh13gardas2016non,nh14matsoukas2023non,nh15bender2007faster,nh16cao2023statistical,nh17giri2009non,nh18ju2019non,nh19ju2024emergent}. In a recent work \cite{Trivedi:2026itv} it was also shown that effective non-Hermiticity globally is strongly constrained by cosmology itself, using an underlying approach of quantum cosmology \cite{qc1bojowald2015quantum,qc2bojowald2008loop,qc3misner1969quantum,qc4wiltshire1996introduction,qc5ashtekar2011loop,qc6hawking1987quantum,qc7gell1996quantum,qc8bojowald2011quantum,qc9vilenkin1995predictions,qc10calcagni2017classical}.
\\
\\
In this broader perspective, non-Hermitian quantum mechanics can be interpreted either as a candidate generalization of fundamental dynamics or as an effective description emerging from reduced or conditioned evolution and in both interpretations, the anti-Hermitian component $\hat \Gamma$ provides a quantitative measure of probability flux and irreversibility. This parametrization makes non-Hermitian frameworks particularly valuable for probing departures from strict unitarity and for connecting microscopic dynamics with experimentally measurable decay rates, amplification factors and spectral shifts.

\section{Quantum thermodynamics and Black Holes}
Quantum thermodynamics is the formulation of thermodynamic principles directly at the level of quantum states and quantum channels \cite{qt1vinjanampathy2016quantum,qt2kosloff2013quantum,qt3deffner2019quantum}. Rather than beginning from macroscopic assumptions such as weak correlations, negligible fluctuations, or a sharply defined split between system and bath, the modern framework takes as fundamental the density operator $\rho$ acting on a Hilbert space $\mathcal{H}$ and the structure of completely positive trace preserving (CPTP) maps that govern its evolution \cite{qt4pekola2015towards,qt5campbell2026roadmap,qt6alicki2019introduction}. The basic thermodynamic quantities are therefore intrinsically quantum and the entropy of a state is given by the von Neumann entropy
\begin{equation}
S(\rho) = -\mathrm{Tr}\big(\rho \ln \rho\big)
\end{equation}
which reduces to the Shannon entropy for classical probability distributions but, in general, encodes quantum coherence and entanglement. Closely related is the relative entropy between two states $\rho$ and $\sigma$ \cite{qt7millen2016perspective,qt8itoi2020second}
\begin{equation}
S(\rho\Vert \sigma) = \mathrm{Tr}\!\left[\rho\big(\ln \rho - \ln \sigma\big)\right]
\end{equation}
which measures distinguishability and plays a central role in quantifying irreversibility. A cornerstone result of quantum information theory is that relative entropy is monotone under CPTP maps, a statement often referred to as the data-processing inequality. This monotonicity under physical evolution provides the quantum analogue of the second law as entropy production can be expressed in terms of the decrease of relative entropy with respect to an appropriate stationary state \cite{qt9itoi2017universal,qt10gemmer2009quantum}. In this sense, quantum thermodynamics replaces phenomenological entropy balance statements by precise inequalities derived from the structural properties of quantum channels.
\\
\\
In addition to entropy, quantum thermodynamics also refines the notions of work, heat and energy exchange. For a system evolving under a time-dependent Hamiltonian $H(t)$ and a CPTP map describing environmental coupling, the change in internal energy $U = \mathrm{Tr}(\rho H)$ can be decomposed into coherent driving contributions and dissipative fluxes. Importantly, entropy production can be identified at the level of the density matrix itself, without assuming proximity to equilibrium and this makes the framework particularly well suited for strongly correlated or highly entangled systems, where classical thermodynamic intuition may fail \cite{qt11binder2015quantum,qt12kammerlander2016coherence,qt13von2014some}. The formalism is therefore inherently microscopic, fully compatible with quantum coherence and capable of treating non-equilibrium processes in a unified manner.
\\
\\
These features become essential in the context of black holes as near a black hole horizon, the assumptions underlying classical thermodynamics break down in several crucial ways. The exterior region is not an isolated system as quantum fields propagate across the horizon and the vacuum state is highly entangled between interior and exterior modes. An observer restricted to the exterior algebra must describe physics using a reduced density matrix obtained by tracing over inaccessible interior degrees of freedom and even if the global state is pure and evolves unitarily, the reduced exterior state generically evolves non-unitarily and is naturally described by an open system dynamics. Dissipation, decoherence and entropy production thus arise not from phenomenological modeling but from the fundamental causal structure of spacetime. This is precisely the regime in which quantum thermodynamics provides the appropriate conceptual and mathematical language since it identifies which quantities remain well defined and monotone under coarse-graining and reduced evolution.
\\
\\
In black hole physics, the relevant thermodynamic quantity is not the entropy of exterior fields alone but the generalized entropy \cite{bh1carlip2014black}. For a spatial slice $\Sigma$ intersecting the horizon on a codimension-two surface, one  can then define
\begin{equation}
S_{\mathrm{gen}}(\Sigma) = \frac{A(\Sigma)}{4G\hbar} + S_{\mathrm{out}}(\Sigma) + S_{\mathrm{ct}}(\Sigma)
\end{equation}
where $A(\Sigma)$ is the area of the horizon cross section, $S_{\mathrm{out}}(\Sigma)$ is the von Neumann entropy of quantum fields restricted to the exterior region and $S_{\mathrm{ct}}(\Sigma)$ denotes local counterterms required to renormalize entanglement divergences. The area term reflects the geometric entropy associated with the horizon, while $S_{\mathrm{out}}$ captures the quantum informational content accessible to the exterior observer. The generalized second law asserts that for physically admissible processes the generalized entropy is non-decreasing,
\begin{equation}
\frac{d}{dt}S_{\mathrm{gen}} \ge 0
\end{equation}
From the standpoint of quantum thermodynamics, this inequality is understood as an entropy balance law for an open quantum system coupled to geometric degrees of freedom. The monotonicity properties of relative entropy, together with the structure of reduced dynamics, ensure that entropy production in the exterior sector is compensated by the appropriate change in horizon entropy. The necessity of quantum thermodynamics near black holes is therefore not optional but dictated by the operational fact that any exterior description is intrinsically open and that the second law must be formulated in terms of $S_{\mathrm{gen}}$, rather than $S_{\mathrm{out}}$ alone.

\section{Inevitability of Effective non-Hermiticity near Black Holes}
Let us now clarify why quantum thermodynamics, black hole physics and non-Hermiticity can naturally converge in a single discussion. Quantum thermodynamics provides the correct microscopic formulation of the second law for open quantum systems, expressing irreversibility in terms of entropy production and the monotonicity of relative entropy under completely positive trace-preserving evolution. Black holes, on the other hand introduce causal horizons that force any exterior observer to adopt a reduced description obtained by tracing over inaccessible degrees of freedom. Such a reduction generically converts globally unitary evolution into locally non-unitary dynamics and non-Hermiticity can then enter not as a speculative modification of quantum theory, but as the inevitable effective structure that encodes probability and information flux into the traced-out sector.  It may therefore be conceptually unavoidable to analyze quantum thermodynamics, horizon physics and effective non-Hermitian generators together as they are different facets of the same structural fact that restricted observers in curved spacetime describe open quantum systems whose dynamics is most naturally represented by a Lindblad generator with an effective non-Hermitian Hamiltonian at the amplitude level.
\\
\\
Now, let's consider a global Hilbert space that factors into exterior and interior degrees of freedom,
\begin{equation}
\mathcal{H}=\mathcal{H}_{\mathrm{out}}\otimes \mathcal{H}_{\mathrm{in}}
\end{equation}
and let $\rho(t)$ be the global density matrix and the exterior state accessible to an asymptotic observer is
\begin{equation}
\rho_{\mathrm{out}}(t)=\mathrm{Tr}_{\mathrm{in}}\big(\rho(t)\big)
\end{equation}
Even if the global state evolves unitarily,
\begin{equation}
\frac{d\rho}{dt}=-i\big[H_{\mathrm{tot}},\rho\big]
\end{equation}
the reduced state $\rho_{\mathrm{out}}$ generally satisfies a non-unitary evolution equation and so under the usual physical requirements for reduced dynamics, namely linearity complete positivity and trace preservation, the most general Markovian form is known as a Lindblad master equation of the form
\begin{equation}
\frac{d\rho_{\mathrm{out}}}{dt} = -i\big[H,\rho_{\mathrm{out}}\big]
+ \sum_a\left(L_a\rho_{\mathrm{out}}L_a^\dagger-\frac12\big\{L_a^\dagger L_a,\rho_{\mathrm{out}}\big\}\right)
\end{equation}
where $H$ is Hermitian and the $L_a$ encode the influence of the traced-out sector. It is algebraically useful to rewrite the generator in a form that makes the effective non-Hermitian Hamiltonian explicit and so we define
\begin{equation}
H_{\mathrm{eff}} = H - \frac{i}{2}\sum_a L_a^\dagger L_a
\end{equation}
and note that one may express the Lindblad equation as
\begin{equation}
\frac{d\rho_{\mathrm{out}}}{dt}
= -i\left(H_{\mathrm{eff}}\rho_{\mathrm{out}}-\rho_{\mathrm{out}}H_{\mathrm{eff}}^\dagger\right) + \sum_a L_a\rho_{\mathrm{out}}L_a^\dagger
\end{equation}
Writing
\begin{equation}
\Gamma \equiv \frac12\sum_a L_a^\dagger L_a
\end{equation}
one has $H_{\mathrm{eff}}=H-i\Gamma$ with $\Gamma\succeq 0$ and note that importantly the appearance of $H-i\Gamma$ is not an ad hoc modification but it is a representation of the most general completely positive reduced dynamics. In particular, a naive non-Hermitian Schr\"odinger equation without the jump terms would not be trace preserving and would not define a consistent probabilistic theory, whereas the Lindblad completion restores trace preservation and ensures physicality.
\\
\\
The connection to the generalized second law is immediate once one recognizes that the exterior sector is open and thus produces entropy and so we let $S_{\mathrm{out}}(t)=S(\rho_{\mathrm{out}}(t))$ and then differentiating the von Neumann entropy gives us
\begin{equation}
\frac{dS_{\mathrm{out}}}{dt}
= -\mathrm{Tr}\!\left(\frac{d\rho_{\mathrm{out}}}{dt}\ln \rho_{\mathrm{out}}\right)
\end{equation}
where $\mathrm{Tr}(d\rho_{\mathrm{out}}/dt)=0$ has been used. The purely Hamiltonian part does not contribute because $\mathrm{Tr}([H,\rho]\ln\rho)=0$, hence only the dissipator contributes to entropy change. A sharper statement is obtained by introducing a stationary reference state $\rho_{\mathrm{ss}}$ satisfying $\mathcal{L}(\rho_{\mathrm{ss}})=0$ for the Lindblad generator $\mathcal{L}$. We now define the entropy production rate by  given by the so called Spohn's formula,
\begin{equation}
\Pi(t)\equiv -\frac{d}{dt}S\!\left(\rho_{\mathrm{out}}(t)\Vert \rho_{\mathrm{ss}}\right)
\end{equation}
Using the definition of relative entropy, one computes
\begin{equation}
\frac{d}{dt}S\!\left(\rho\Vert \rho_{\mathrm{ss}}\right)
= \mathrm{Tr}\!\left(\frac{d\rho}{dt}\ln\rho\right)
- \mathrm{Tr}\!\left(\frac{d\rho}{dt}\ln\rho_{\mathrm{ss}}\right)
\end{equation}
hence
\begin{equation}
\Pi(t)
=  -\mathrm{Tr}\!\left(\frac{d\rho}{dt}\ln\rho\right)
+ \mathrm{Tr}\!\left(\frac{d\rho}{dt}\ln\rho_{\mathrm{ss}}\right)
=\frac{dS}{dt} + \mathrm{Tr}\!\left(\mathcal{L}(\rho)\ln\rho_{\mathrm{ss}}\right)
\end{equation}
Monotonicity of relative entropy under completely positive trace preserving maps implies $\Pi(t)\ge 0$. This inequality is the quantum second law in a form appropriate for open systems and it is essential that this statement depends on complete positivity and trace preservation, not on the Hermiticity of an effective Hamiltonian. The effective non-Hermiticity $H-i\Gamma$ is fully consistent with $\Pi\ge 0$ because the physical evolution is defined at the level of $\rho$ by a CPTP generator.
\\
\\
Now comes the beautiful part, where we connect this entropy balance to black hole thermodynamics. For this we consider a stationary black hole with Hawking temperature
\begin{equation}
T_H = \frac{\hbar\kappa}{2\pi}
\end{equation}
with $\kappa$ the surface gravity and for near-horizon physics it is natural to take $\rho_{\mathrm{ss}}$ to be a KMS-like stationary state at inverse temperature $\beta_H=1/T_H$ for the exterior algebra. If $\rho_{\mathrm{ss}}\propto e^{-\beta_H H}$ in an effective description, then
\begin{equation}
\ln \rho_{\mathrm{ss}} = -\beta_H H - \ln Z
\end{equation}
and since $\mathrm{Tr}(\mathcal{L}(\rho))=0$ the constant $-\ln Z$ drops out and one obtains
\begin{equation}
\Pi = \frac{dS_{\mathrm{out}}}{dt} - \beta_H \frac{d}{dt}\langle H\rangle_{\mathrm{out}}
\end{equation}
where $\langle H\rangle_{\mathrm{out}}=\mathrm{Tr}(\rho_{\mathrm{out}}H)$. Rearranging this gives us a Clausius type inequality which is of the form
\begin{equation}
\frac{dS_{\mathrm{out}}}{dt} \ge \beta_H \frac{d}{dt}\langle H\rangle_{\mathrm{out}}
\end{equation}
The right hand side is an energy flux term very importantly. In black hole physics energy flux across the horizon is tied to the change in horizon area via the first law of black hole mechanics and for a quasi-stationary process,
\begin{equation}
dM = \frac{\kappa}{8\pi G}dA + \cdots
\end{equation}
where the ellipsis denotes work terms such as $\Omega dJ$ and $\Phi dQ$ that may be included when appropriate. For simplicity, focusing on the energy exchange term, the area change satisfies
\begin{equation}
dA = \frac{8\pi G}{\kappa}dM
\end{equation}
Multiplying by $1/(4G\hbar)$ gives
\begin{equation}
d\left(\frac{A}{4G\hbar}\right) = \frac{2\pi}{\hbar\kappa}dM = \beta_H dM
\end{equation}
At the level of rates, writing $dM/dt$ as the net energy flux into the black hole one has
\begin{equation}
\frac{d}{dt}\left(\frac{A}{4G\hbar}\right) = \beta_H \frac{dM}{dt}
\end{equation}
Energy conservation between exterior fields and the black hole implies that the exterior energy change satisfies
\begin{equation}
\frac{d}{dt}\langle H\rangle_{\mathrm{out}} = -\frac{dM}{dt}
\end{equation}
with sign conventions fixed so that $dM/dt>0$ corresponds to energy entering the hole and then combining these relations ends up giving us
\begin{equation}
\beta_H \frac{d}{dt}\langle H\rangle_{\mathrm{out}} = -\frac{d}{dt}\left(\frac{A}{4G\hbar}\right)
\end{equation}
Substituting into the Clausius inequality gives
\begin{equation}
\frac{dS_{\mathrm{out}}}{dt} \ge -\frac{d}{dt}\left(\frac{A}{4G\hbar}\right)
\end{equation}
or equivalently
\begin{equation}
\frac{d}{dt}\left(\frac{A}{4G\hbar}+S_{\mathrm{out}}\right)\ge 0
\end{equation}
This is precisely the generalized second law in its simplest semiclassical form. The crucial observation is that if the exterior dynamics were strictly unitary, $dS_{\mathrm{out}}/dt$ would vanish, and processes with $dA/dt<0$ would violate the generalized second law. The generalized second law therefore requires that the exterior description include irreversible dynamics, which in turn is most naturally encoded by an effective non-Hermitian generator $H-i\Gamma$ at the amplitude level together with the corresponding jump terms at the density matrix level. So, for thermodynamics to really work near black holes, non-Hermiticity becomes non-negotiable.
\\
\\
The emphasis in this derivation is that the quantum laws used here, including monotonicity of relative entropy and Spohn's inequality, are statements about completely positive trace preserving evolution of $\rho_{\mathrm{out}}$. They do not assume that an effective Hamiltonian $H_{\mathrm{eff}}$ is Hermitian and in fact, the effective non-Hermiticity is part of the general structure of open quantum system dynamics. The fundamental requirement is not Hermiticity but the physicality of the reduced dynamics as a quantum channel.
\\
\section{Hermiticity as a Symmetry Law}
The discussion so far has deliberately used black holes as more than a dramatic example of open quantum dynamics. They provide a clean and unavoidable arena in which global unitarity and local non-unitarity coexist in a sharply defined way. In ordinary laboratory systems, non-Hermiticity is often introduced phenomenologically to model gain, loss or environmental coupling but in contrast, black hole spacetimes force an observer-dependent reduction of the Hilbert space through causal structure alone with the horizons partition of the global state into accessible and inaccessible sectors, and tracing over the latter generically produces non-unitary evolution in the former. The generalized second law then appears not as an abstract thermodynamic principle, but as a precise entropy balance relation for this reduced quantum system, where geometric entropy compensates for entropy production in the exterior algebra. This raises a deeper structural question, which is that if effective non-Hermiticity is inevitable in the presence of horizons and if horizons are intrinsic features of gravitational dynamics, then perhaps the relation between gravity and non-Hermiticity is not accidental but structural? In that case, the familiar Hermiticity condition of quantum mechanics might itself be understood not as a primitive postulate, but as the manifestation of a more fundamental symmetry which is namely, the global conservation of an inner product charge that holds only when the spacetime admits complete Cauchy surfaces without boundary flux. Black holes then serve as the natural counterexample where they expose precisely how and why that conservation law can fail for restricted observers. The purpose of the present section is to elevate this observation into a principle where we reinterpret Hermiticity as a symmetry statement tied to the existence of a globally conserved inner product current. When this symmetry is exact, standard unitary quantum mechanics is recovered but when it is obstructed by horizons or causal boundaries, effective non-Hermiticity emerges in the reduced description. In this way, gravity, non-Hermitian dynamics and entropy balance are unified under a single structural theme, with Hermiticity appearing as the symmetry that connects them.
\\
\\
One may now re-express the origin of Hermiticity as a symmetry statement about inner products as well. In standard quantum mechanics, the conservation of probabilities may be viewed as the conservation of the norm, which may be generalized by introducing a positive metric operator $\eta$ defining an inner product
\begin{equation}
\langle \psi|\phi\rangle_\eta = \langle \psi|\eta|\phi\rangle
\end{equation}
Norm conservation is the requirement
\begin{equation}
\frac{d}{dt}\langle \psi|\eta|\psi\rangle = 0
\end{equation}
Using the Schr\"odinger equation $i\partial_t|\psi\rangle = H|\psi\rangle$ and its adjoint, one finds
\begin{equation}
\frac{d}{dt}\langle \psi|\eta|\psi\rangle
= i\langle \psi|H^\dagger\eta-\eta H|\psi\rangle
+ \langle \psi|\dot{\eta}|\psi\rangle
\end{equation}
Requiring this to vanish for all $|\psi\rangle$ gives us the operator identity
\begin{equation}
H^\dagger\eta-\eta H = i\dot{\eta}
\end{equation}
In flat space with the canonical inner product, $\eta=\mathbb{I}$ and $\dot{\eta}=0$, this reduces to
\begin{equation}
H^\dagger = H
\end{equation}
so standard Hermiticity is recovered as a special case but in more general situations, $\eta$ can encode background structure including effective metric dependence and a time-dependent $\eta$ corresponds to the failure of strict Hermiticity in the naive inner product. From this viewpoint, Hermiticity is a manifestation of a conserved "inner product charge".
\\
\\
To express this covariantly, we can consider a relativistic quantum field and its associated conserved current in a globally hyperbolic region. For example, for a complex Klein-Gordon field $\Phi$ on a curved background with metric $g_{\mu\nu}$ one defines
\begin{equation}
j^\mu = i\left(\Phi^\ast \nabla^\mu \Phi - \Phi \nabla^\mu \Phi^\ast\right)
\end{equation}
and on solutions of the field equations one has
\begin{equation}
\nabla_\mu j^\mu = 0
\end{equation}
The conserved inner product on a Cauchy surface $\Sigma$ with future-directed unit normal $n_\mu$ is
\begin{equation}
(\Phi_1,\Phi_2)=\int_\Sigma d\Sigma\, n_\mu\, j^\mu(\Phi_1,\Phi_2)
\end{equation}
with $d\Sigma$ the induced volume element. Conservation of the inner product is then the statement that the value of this integral is independent of the choice of Cauchy surface, which follows from $\nabla_\mu j^\mu=0$ and Stokes's theorem in a globally hyperbolic region without flux through boundaries. In the presence of boundaries such as horizons, the conservation law becomes an exact flux balance instead of an equality between two Cauchy surfaces and if $\mathcal{R}$ is a spacetime region bounded by two hypersurfaces $\Sigma_1$ and $\Sigma_2$ and a boundary $\mathcal{B}$, then
\begin{equation}
\int_{\Sigma_2} d\Sigma\, n_\mu j^\mu - \int_{\Sigma_1} d\Sigma\, n_\mu j^\mu = -\int_{\mathcal{B}} d\Sigma_\mu\, j^\mu
\end{equation}
For an exterior observer in a black hole spacetime, $\mathcal{B}$ includes the horizon and the flux term is generically nonzero. The inner product "charge" in the exterior region is therefore not conserved even if $\nabla_\mu j^\mu=0$ holds locally, because conservation reduces to a statement with boundary flux. This is the covariant sense in which the inner product current is obstructed for restricted observers.
\\
\\
\textbf{Hermiticity in quantum mechanics is thus seen here as a manifestation of a global conservation law for the inner product current. In spacetimes with causal horizons, this conservation law is obstructed for restricted observers leading to effective non-Hermitian dynamics and entropy production. The generalized second law is the statement that the geometric entropy of the horizon compensates for the non-conservation of the inner product charge.} This construction suggests that Hermiticity is best viewed as a statement about the existence of a globally conserved inner product, which in covariant language is tied to the existence of a conserved current and to the ability to define global Cauchy surfaces without boundary flux. A concise way to encode this is to define the inner product charge on a hypersurface $\Sigma$ by
\begin{equation}
Q[\Sigma]=\int_{\Sigma} d\Sigma\, n_\mu J^\mu
\end{equation}
where $J^\mu$ is the appropriate inner product current for the matter system under consideration and so in globally hyperbolic spacetimes without horizons and with suitable boundary conditions at infinity, one has
\begin{equation}
Q[\Sigma_2]=Q[\Sigma_1]
\end{equation}
for any two Cauchy surfaces $\Sigma_1$ and $\Sigma_2$. In such settings, one may choose a time foliation and interpret $Q$ as the conserved norm, thus recovering the standard probabilistic interpretation. In particular, when the background is exactly Minkowski and the foliation is standard, the conserved inner product becomes time-independent and the generator of time translations may be taken Hermitian in the canonical inner product.
\\
\\
In the presence of horizons or other causal boundaries the correct statement is
\begin{equation}
Q[\Sigma_2]-Q[\Sigma_1] = -\int_{\mathcal{B}} d\Sigma_\mu\, J^\mu
\end{equation}
so the failure of conservation in a restricted region is precisely equal to the flux through the boundary. This is not a microscopic violation of probability conservation but an accounting statement which is that the charge is transported into degrees of freedom that are not part of the reduced description. The effective non-Hermiticity of the reduced theory is a compact way to represent this flux when one describes the exterior sector alone. One may also view the curved background as entering the quantum theory through the definition of inner products and mode decomposition as the covariant conservation law
\begin{equation}
\nabla_\mu J^\mu = 0
\end{equation}
is itself metric dependent because $\nabla_\mu$ is the Levi-Civita derivative associated with $g_{\mu\nu}$. Furthermore we can see that the hypersurface element $d\Sigma\, n_\mu$ depends on the induced metric and the unit normal and so the numerical value of $Q[\Sigma]$ is computed by combining matter data with geometric data. This is a precise sense in which the metric properties are encoded in the basic Hilbert space structure of quantum theory in curved spacetime as when the metric becomes flat, $\nabla_\mu\to \partial_\mu$ and the hypersurface measure reduces to the standard one, giving us the conventional conserved norm and the standard Hermitian generator for time evolution.
\\
\\
A complementary perspective uses the generalized inner product with a metric operator $\eta$ on the Hilbert space. In curved space or time-dependent backgrounds, the natural inner product relevant for an observer adapted foliation may be written in the form
\begin{equation}
\langle \psi|\phi\rangle_\eta = \langle \psi|\eta(g_{\mu\nu})|\phi\rangle
\end{equation}
with $\eta$ depending functionally on the spacetime geometry and the choice of slicing and the condition for norm conservation,
\begin{equation}
H^\dagger\eta-\eta H = i\dot{\eta}
\end{equation}
then shows that even if the fundamental dynamics is unitary in a covariant sense, a time-dependent effective metric operator may appear in a reduced descriptions, producing apparent non-Hermiticity in the canonical inner product. In the strict flat-space limit, $\eta\to \mathbb{I}$ and $\dot{\eta}\to 0$, recovering ordinary Hermiticity and the standard probability interpretation.
\\
\\
But what does it all exactly mean physically? The non conservation of the inner product charge in a restricted region has a clear operational meaning in this regard. The full theory may possess a globally conserved current $J^\mu$ and a globally unitary time evolution, but an observer who only has access to a subalgebra of observables must describe the system by a reduced density matrix. The reduced dynamics is then generically non-unitary and the apparent breakdown of norm conservation or Hermiticity is not a fundamental failure of quantum mechanics but a reflection of tracing over inaccessible degrees of freedom. In this sense foundational probability may be conserved in the UV complete theory while the effective theory for accessible observables is not.
\\
\\
The statement that gravity "breaks" probability is therefore misleading as well and we need to be precise on what is really happening. What gravity introduces through horizons and causal structure, is a sharp notion of inaccessible regions and boundary flux. The relevant conservation law becomes a balance equation involving horizon flux terms and in practice, horizons plus coarse-graining break our ability to track the conserved charge within the exterior sector alone. The open-system description is the correct operational description and the effective non-Hermiticity encoded by $H_{\mathrm{eff}}=H-i\Gamma$ is a bookkeeping device for probability and information flow into the traced-out sector.
\\
\\
This viewpoint illuminates aspects of the information paradox as well. If the global evolution is unitary, information is not destroyed as it is redistributed into correlations between exterior and interior degrees of freedom or more generally into degrees of freedom not captured by a naive semiclassical exterior effective field theory. The reduced exterior description exhibits entropy production and information loss because it discards correlations with the inaccessible sector. The generalized entropy provides precisely the correct accounting variable which is that the horizon area term compensates for the loss of trackable information in the exterior sector, so that the generalized second law holds. Any resolution of the information paradox must therefore specify how the bookkeeping changes when quantum gravity effects are included, whether by allowing information to be recovered in late-time radiation, by modifying the semiclassical factorization or by introducing additional degrees of freedom associated with horizon microphysics. The central lesson is that effective non-unitarity in the exterior sector is not itself paradoxical but it is the expected operational signature of horizons and coarse-graining.
\\
\section{Hermiticity and General Relativity} Up to this point, we have argued that effective non-Hermiticity emerges inevitably in the presence of horizons because restricted observers must describe an open quantum system. We have further suggested that Hermiticity itself can be reinterpreted as the global conservation of an inner-product charge and if this reinterpretation is correct, then the conditions under which Hermiticity holds or fails should be controlled not merely by quantum kinematics, but by spacetime geometry. General relativity is precisely the framework that determines causal structure, global hyperbolicity, horizon formation and flux balance laws. Its field equations encode local conservation identities and govern how energy flux reshapes geometry and it is therefore natural to ask whether the very structure of Einstein’s equations provides the geometric consistency conditions required for global inner-product conservation and whether modifications of gravitational dynamics correspondingly modify the circumstances under which Hermiticity can be globally implemented. In this section, we explore this possibility in detail, showing how conservation laws arising from the Bianchi identity, horizon focusing equations and geometric entropy functionals collectively determine the conditions under which Hermitian quantum evolution can be consistently realized for restricted observers.
\\
\\
A profound structural link between gravity and conservation laws is already encoded at the purely geometric level. In any pseudo-Riemannian manifold endowed with Levi-Civita connection $\nabla_\mu$, the Einstein tensor
\begin{equation}
G_{\mu\nu} = R_{\mu\nu} - \frac12 R g_{\mu\nu}
\end{equation}
satisfies the contracted Bianchi identity
\begin{equation}
\nabla_\mu G^{\mu\nu} = 0
\end{equation}
This identity is not dynamical but geometric as it follows from the symmetries of the Riemann tensor and the definition of $G_{\mu\nu}$. If spacetime dynamics is governed by Einstein’s equations
\begin{equation}
G^{\mu\nu} + \Lambda g^{\mu\nu} = 8\pi G\, T^{\mu\nu}
\end{equation}
then taking the covariant divergence of both sides and using $\nabla_\mu g^{\mu\nu}=0$ immediately yields
\begin{equation}
\nabla_\mu T^{\mu\nu} = 0
\end{equation}
Thus the local covariant conservation of stress-energy is not an independent postulate but a consistency condition enforced by spacetime geometry itself. In local inertial coordinates this reduces to $\partial_\mu T^{\mu\nu}=0$, which is the familiar statement of energy-momentum conservation. For quantum fields, $T^{\mu\nu}$ is an operator valued distribution and its conservation must be understood after suitable renormalization but nevertheless, the geometric structure compels the matter sector to respect a compatible conservation law if the coupled theory is to be consistent.
\\
\\
This observation parallels the inner product current conservation discussed previously as there, the conserved current
\begin{equation*}
\nabla_\mu J^\mu = 0
\end{equation*}
expressed the local statement underlying global norm conservation. The existence of a time-independent inner product charge
\begin{equation*}
Q[\Sigma] = \int_\Sigma d\Sigma\, n_\mu J^\mu
\end{equation*}
requires not only the local conservation law but also suitable global conditions which are global hyperbolicity, the absence of boundary flux, and the existence of well-defined Cauchy surfaces. These are precisely the same geometric conditions that guarantee a well-posed Cauchy problem for matter fields in curved spacetime. In this sense, the gravitational field equations and their associated conservation identities provide the geometric infrastructure within which the quantum probabilistic structure can be globally implemented. Hermiticity, interpreted as conservation of an inner-product charge, is therefore meaningful only when spacetime geometry allows that charge to be globally conserved.
\\
\\
The connection becomes sharper when one considers black hole horizons and the dynamics of null congruences. For seeing this, let $k^\mu$ denote the tangent vector to null generators of a horizon and $\lambda$ an affine parameter along them and then the Raychaudhuri equation for the expansion $\theta = \nabla_\mu k^\mu$ reads
\begin{equation}
\frac{d\theta}{d\lambda}
= -\frac12 \theta^2 - \sigma_{\mu\nu}\sigma^{\mu\nu}
- R_{\mu\nu}k^\mu k^\nu
\end{equation}
where $\sigma_{\mu\nu}$ is the shear tensor. For a quasi-stationary horizon, $\theta$ and $\sigma_{\mu\nu}$ are small and the dominant source term arises from the Ricci tensor contraction. Using Einstein’s equations, one writes
\begin{equation}
R_{\mu\nu}k^\mu k^\nu
= 8\pi G\, T_{\mu\nu}k^\mu k^\nu
\end{equation}
since $g_{\mu\nu}k^\mu k^\nu=0$ for null vectors eliminates the trace term and thus matter energy flux through the horizon directly controls the focusing of null generators. The expansion determines the change in cross-sectional area $A$ of the congruence via
\begin{equation}
\frac{dA}{d\lambda} = \theta A
\end{equation}
Combining these relations yields, after integrating along the generators,
\begin{equation}
\frac{d^2 A}{d\lambda^2}
\sim -8\pi G\, A\, T_{\mu\nu}k^\mu k^\nu
\end{equation}
demonstrating explicitly that stress-energy flux governs horizon area change. Through the identification of horizon entropy with $A/4G\hbar$, this geometric response becomes the entropy variation required to balance exterior entropy production in the generalized second law. The Einstein equations therefore supply precisely the geometric response mechanism needed to compensate for the effective openness of the exterior quantum system as the conservation law $\nabla_\mu T^{\mu\nu}=0$ ensures consistency of energy accounting, while the Raychaudhuri equation translates flux into area change, closing the thermodynamic balance.
\\
\\
This structural harmony between Einstein dynamics and entropy balance suggests a deeper link. If Hermiticity corresponds to global conservation of an inner-product charge, then the ability to define such a conserved charge depends on the existence of complete Cauchy surfaces without boundary flux and the horizons obstruct this conservation by introducing a boundary through which the current can flow. In such cases, one has
\begin{equation}
Q[\Sigma_2] - Q[\Sigma_1]
= -\int_{\mathcal B} d\Sigma_\mu J^\mu
\end{equation}
with $\mathcal B$ including the horizon. The geometric dynamics of the horizon governed by Einstein’s equations then determine how this flux is encoded in area change. The generalized second law becomes the global balance statement
\begin{equation}
\frac{d}{dt}\left(\frac{A}{4G\hbar} + S_{\mathrm{out}}\right) \ge 0
\end{equation}
which may be interpreted as the statement that geometric entropy compensates for the non-conservation of the inner-product charge in the exterior region. In this sense, the specific structure of Einstein’s equations underwrites the simplest realization of Hermiticity as a global symmetry law.
\\
\\
If one modifies the gravitational action, this balance generically changes as well , for this consider an effective action of the form
\begin{equation}
S_{\mathrm{grav}}
= \frac{1}{16\pi G}
\int d^4x\, \sqrt{-g}
\left(
R + \alpha R^2 + \beta R_{\mu\nu}R^{\mu\nu} + \cdots
\right)
\end{equation}
Varying this action yields modified field equations containing higher derivative terms. The geometric entropy associated with stationary horizons is no longer simply proportional to area but is given by the Wald entropy functional,
\begin{equation}
S_{\mathrm{grav}}
= -2\pi
\int_\Sigma d^2x\, \sqrt{h}\,
\frac{\partial \mathcal L}{\partial R_{\mu\nu\rho\sigma}}
\epsilon_{\mu\nu}\epsilon_{\rho\sigma}
\end{equation}
where $\mathcal L$ is the Lagrangian density and then the generalized entropy becomes
\begin{equation}
S_{\mathrm{gen}} = S_{\mathrm{grav}} + S_{\mathrm{out}} + S_{\mathrm{ct}}
\end{equation}
Because the geometric response law is modified, the precise relation between energy flux, focusing and entropy variation also changes. The compensation mechanism between horizon entropy and exterior entropy production becomes theory dependent and then in an effective description, this can alter the extent to which the exterior sector admits a time-independent inner product and a strictly Hermitian effective generator. Loop corrections and quantum vacuum effects which motivate higher-curvature terms, therefore potentially introduce richer structures in the interplay between geometry and effective non-Hermiticity.
\\
\\
These considerations extend beyond classical general relativity generically as well. In quantum field theory on curved spacetime, the probabilistic structure is implemented locally but global notions of vacuum, particle content and unitarity depend sensitively on geometry and observer access. Horizons naturally render the exterior theory open, leading to Lindblad-type evolution, entropy production and effective non-Hermitian generators. In more general high energy frameworks incorporating nonlocality, higher derivatives or additional degrees of freedom, the open system and thermodynamic description of accessible sectors becomes even more central. The appropriate language in such regimes is not strict subsystem unitarity but that of quantum channels, generalized entropies and covariant flux balances.
\\
\section{Observational Probes from Black Hole Ringdown and Spectroscopy}

If Hermiticity in the canonical inner product is reinterpreted as the special case of a globally conserved inner-product current and if horizons obstruct this conservation through a boundary flux, then the framework developed in this work admits observational consequences in principle. In the presence of a causal horizon, the exterior sector is described by a reduced open quantum system and the effective dynamics at the amplitude level is governed by a non-Hermitian generator as in \eqref{eq:H_decomp} with $\Gamma > 0$ encoding the inner-product flux into the inaccessible sector. To render such effects testable, we introduce a small dimensionless parameter $\varepsilon$ characterizing the strength of the horizon-induced inner-product flux in a given perturbative sector. In the limit $\varepsilon \to 0$ one recovers standard Hermitian evolution and classical general relativity, while nonzero $\varepsilon$ parametrizes controlled departures associated with effective non-Hermiticity for restricted observers. \\

We now consider how the natural arena in which such effects can manifest is black hole ringdown \cite{qnm1nollert1996significance,qnm2berti2009quasinormal,qnm8giesler2019black,qnm7isi2021analyzing}. Linear perturbations of a stationary black hole spacetime reduce to master equations of the Regge–Wheeler or Zerilli type and there, after separation of variables
\begin{equation}
\Psi(t,r,\theta,\phi) = e^{-i\omega t} Y_{\ell m}(\theta,\phi)\, \psi(r)
\end{equation}
the radial function satisfies
\begin{equation}
\left[ \frac{d^2}{dr_*^2} + \omega^2 - V_\ell(r) \right] \psi(r) = 0 
\end{equation}
where $r_*$ is the tortoise coordinate and $V_\ell(r)$ is the effective potential determined by the background geometry. In classical general relativity, quasi-normal modes are defined by the boundary conditions \cite{qnm3konoplya2011quasinormal,qnm4baibhav2018black,qnm5mitman2023nonlinearities,qnm6silva2023black}
\begin{equation}
\psi(r_*\to -\infty) \sim e^{-i\omega r_*}, 
\qquad
\psi(r_*\to +\infty) \sim e^{+i\omega r_*}
\end{equation}
corresponding to purely ingoing waves at the horizon and purely outgoing waves at spatial infinity. These boundary conditions render the spectral problem non-selfadjoint producing a discrete set of complex frequencies
\begin{equation}
\omega_n^{(0)} = \omega_{R,n}^{(0)} + i \omega_{I,n}^{(0)}, 
\qquad \omega_{I,n}^{(0)} < 0
\end{equation}
In the present framework, we note crucially that the horizon is not merely an absorber of classical energy but a boundary through which inner-product charge flows. The obstruction of global inner-product conservation in the exterior region implies that the effective boundary condition at the horizon can receive a small correction proportional to $\varepsilon$ and the minimal parametrization of this effect is to allow a small admixture of the outgoing mode at the horizon,
\begin{equation}
\psi(r_*\to -\infty) = e^{-i\omega r_*} + \varepsilon\, \mathcal{R}_H(\omega)\, e^{+i\omega r_*}
\end{equation}
where $\mathcal{R}_H(\omega)$ is a dimensionless function determined by the microscopic mechanism of flux transfer and when $\varepsilon = 0$ one recovers the classical purely ingoing condition. The quasinormal frequencies are determined by a spectral condition that may be written abstractly as
\begin{equation}
F(\omega,\varepsilon) = 0 
\end{equation}
Expanding perturbatively then in $\varepsilon$,
\begin{equation}
F(\omega,\varepsilon) = F(\omega,0) + \varepsilon \left( \frac{\partial F}{\partial \varepsilon} \right)_{\varepsilon=0} + \mathcal{O}(\varepsilon^2)
\end{equation}
and writing $\omega_n = \omega_n^{(0)} + \delta \omega_n$, we obtain at linear order
\begin{equation}
\delta \omega_n = - \varepsilon 
\frac{\left( \partial_\varepsilon F \right)_{\omega=\omega_n^{(0)},\,\varepsilon=0}}
{\left( \partial_\omega F \right)_{\omega=\omega_n^{(0)},\,\varepsilon=0}}
\end{equation}
Thus the fractional shifts in the real and imaginary parts of the frequency scale linearly with $\varepsilon$
\begin{equation}
\frac{\Delta \omega_{R,n}}{\omega_{R,n}^{(0)}} \sim \mathcal{O}(\varepsilon),
\qquad
\frac{\Delta \omega_{I,n}}{\omega_{I,n}^{(0)}} \sim \mathcal{O}(\varepsilon)
\end{equation}
Observationally, the ringdown signal is characterized by the frequency
\begin{equation}
f_n = \frac{\omega_{R,n}}{2\pi}
\end{equation}
and damping time
\begin{equation}
\tau_n = \frac{1}{|\omega_{I,n}|}
\end{equation}
A nonzero $\varepsilon$ therefore induces correlated shifts
\begin{equation}
\frac{\Delta f_n}{f_n} \simeq \frac{\Delta \omega_{R,n}}{\omega_{R,n}^{(0)}}
\qquad
\frac{\Delta \tau_n}{\tau_n} \simeq - \frac{\Delta \omega_{I,n}}{\omega_{I,n}^{(0)}}
\end{equation}
These shifts are, in principle, directly measurable in gravitational wave observations of the post merger ringdown phase. Current LIGO–Virgo–KAGRA (LVK) analyses of high signal-to-noise merger events show that the observed dominant quasi-normal modes are consistent with the Kerr predictions of general relativity within present uncertainties, which are typically at the level of $\mathcal{O}(10\%)$ for well measured modes \cite{lvk1LIGOScientific:2025wao,lvk2LIGOScientific:2020tif,lvk3Ghosh:2021mrv,lvk4Pompili:2025cdc}. Interpreting these bounds within the parametrization above and assuming the natural scaling $\Delta \omega / \omega \sim \varepsilon$, we obtain the order of magnitude constraint
\begin{equation}
|\varepsilon| \lesssim 10^{-1}
\end{equation}
in the gravitational perturbation sector currently accessible and note that this does not exclude the present framework, instead it indicates that any horizon induced obstruction of inner-product conservation must be parametrically small for astrophysical black holes probed so far. The limit $\varepsilon \to 0$ remains fully compatible with data, but small nonzero values are not ruled out at present precision.
\\
\\
Future detectors will dramatically improve sensitivity as third generation ground based observatories and space-based missions will measure multiple quasinormal modes with high accuracy, reducing fractional uncertainties to the percent level or below \cite{ngen1Bhagwat:2023jwv,ngen2Baibhav:2018rfk,ngen3Bhagwat:2021kwv,ngen4Maselli:2023khq}. If $\Delta \omega / \omega$ can be constrained at the $\mathcal{O}(10^{-2})$ level, the corresponding bound on $\varepsilon$ tightens to
\begin{equation}
|\varepsilon| \lesssim 10^{-2}
\end{equation}
or smaller, depending on signal-to-noise and mode content. Because the shifts predicted here arise specifically from a modification of the horizon boundary condition, they produce a characteristic correlated pattern across overtones and angular indices, providing a clean target for black hole spectroscopy.
\\
\\
We note that in the calculation of the effect of horizon-induced inner-product flux on black-hole ringdown, one may either retain the standard (real) bulk perturbation equations of general relativity and encode non-Hermiticity through a modified, effectively complex boundary condition at the horizon, or instead introduce a small non-Hermitian deformation of the radial master operator localized near the horizon. Although these two descriptions are equivalent at leading order for computing shifts in quasi-normal mode frequencies, they differ conceptually. The boundary-condition approach attributes the obstruction of Hermiticity to a global flux through a causal boundary, thereby preserving the real bulk geometric dynamics and interpreting non-Hermiticity as a symmetry breaking tied to spacetime structure. By contrast, the bulk-deformation picture treats non-Hermiticity as a local modification of the effective evolution generator in the reduced exterior sector. In the present framework, where Hermiticity arises as a covariant conservation law for an inner-product current, the boundary formulation used above offers the more faithful representation of the underlying geometric symmetry principle, while the bulk non-Hermitian term serves primarily as a calculational device that captures the same horizon-flux physics.
\\
\\
From the perspective developed in this work, black hole ringdown is therefore not merely a test of classical general relativity but a probe of the global conservation of inner-product charge in the presence of horizons. The agreement of current data with Kerr ringdown implies that Hermiticity, in the canonical inner product, is an excellent approximation for the accessible exterior algebra in presently observed systems but at the same time, the absence of detected deviations does not trivialize the framework and instead it translates directly into quantitative upper bounds on the strength of horizon-induced inner-product flux. As observational precision improves over the coming years, black hole spectroscopy will either further tighten these bounds or uncover statistically significant departures from the Kerr spectrum. In either case the idea that Hermiticity is a geometry conditional symmetry law becomes empirically testable, placing it within the domain of precision gravitational wave astronomy.
\\
\\
\section{Conclusions}

In this work we have proposed a structural reinterpretation of Hermiticity as a symmetry law associated with the global conservation of an inner-product current. Beginning from a review of non-Hermitian quantum mechanics and its effective role in open-system dynamics, we emphasized that non-Hermiticity naturally arises whenever a reduced description is adopted. We then showed that black hole spacetimes provide a conceptually unavoidable setting in which such a reduction is enforced by causal structure itself: tracing over interior degrees of freedom renders the exterior algebra effectively open and generically non-unitary. Using the framework of quantum thermodynamics and the monotonicity of relative entropy under completely positive trace-preserving maps, we demonstrated that entropy production in the exterior sector combines with the geometric response of the horizon to yield the generalized second law. This entropy balance can be interpreted as a flux statement for an inner-product charge, with the horizon acting as a boundary through which this charge flows. Hermiticity in the canonical inner product then emerges as the special case in which this flux vanishes globally, a condition that depends on the geometric and causal structure of spacetime. By examining the Bianchi identity, stress-energy conservation, the Raychaudhuri equation and horizon entropy functionals, we established that the specific structure of Einstein’s equations provides the simplest geometric realization of this symmetry and its controlled obstruction in the presence of horizons.
\\
\\
The implications of this perspective are potentially far reaching and that is so for multiple reasons. If Hermiticity is not a primitive postulate but rather the manifestation of a global conservation law tied to spacetime geometry, then gravity plays a fundamental role in determining when and how quantum probabilistic structure can be globally implemented. Horizons, causal boundaries and modifications of gravitational dynamics may alter the conditions under which inner-product conservation holds, thereby influencing the effective emergence of non-Hermitian behavior in restricted sectors. This viewpoint offers a unified language connecting open quantum systems, black hole thermodynamics, information flow and geometric conservation laws and it suggests that the generalized second law may be understood as the organizing principle ensuring consistency between quantum information dynamics and spacetime structure. In a forthcoming sequel work, we will explore the consequences of this framework for quantum field theory in curved spacetime, examining how covariant inner-product conservation, effective non-Hermiticity and horizon flux balance shape the structure of propagators, Green’s functions and effective actions in gravitational backgrounds.
\\
\\
It is also important to note that in the existing literature, the most common way to “reinterpret” Hermiticity is to replace the canonical inner product by a fixed metric operator $\eta$ so that a non-Hermitian $H$ becomes Hermitian in the $\eta$-inner product, $H^\dagger=\eta H\eta^{-1}$, which is pseudo-Hermiticity or in closely related symmetry-based frameworks such as ${\cal PT}$-symmetric quantum mechanics. These approaches primarily address spectral reality and unitary time evolution after an appropriate choice of Hilbert space geometry, but they do not identify why the canonical Hermiticity condition should fail for restricted observers in curved spacetime nor do they tie that failure to a covariant flux law \cite{ph1Mostafazadeh:2001jk,ph2Mostafazadeh:2008pw}. Separately, the open quantum systems literature shows that tracing out inaccessible degrees of freedom generically yields completely positive trace-preserving dynamics and that non-Hermitian effective generators arise naturally at the amplitude level as bookkeeping devices completed by quantum jumps \cite{oq1Lindblad:1975ef,oq2K_nenberg_2017}. The novelty of our present work is to unify all this by elevating Hermiticity to a covariant inner-product symmetry wherein Hermiticity in the canonical inner product is reinterpreted as the special case of global conservation of an inner-product current on Cauchy slices, while black hole horizons obstruct that conservation through a boundary flux, forcing an effective openness of the exterior algebra and hence an inevitable appearance of effective non-Hermiticity for the reduced exterior description. In this view, non-Hermiticity is not introduced ad hoc but it is geometrically mandated by causal boundaries and the generalized second law is re-read as the compensating entropy balance in which horizon entropy precisely accounts for the inner-product charge flux.
\\
\\
\section*{Acknowledgements}
We gratefully acknowledge support from Vanderbilt University and the U.S. National Science Foundation. The work of OT is supported in part by the Vanderbilt Discovery Doctoral Fellowship. The work of AG is supported in part by NSF Award PHY-2411502.

\bibliography{references}
\bibliographystyle{unsrt}

\end{document}